\documentclass[12pt]{article}
\thicklines
\newcommand{\EQ}{\begin{equation}}
\newcommand{\EN}{\end{equation}}
\newcommand{\bea}{\begin{eqnarray}}
\newcommand{\ena}{\end{eqnarray}}
\newcommand{\eea}{\end{eqnarray}}

\def\3i{\int\!\!\!\int\!\!\!\int}
\def\2i{\int\!\!\!\int}

\def\del{\Delta}
\def\ddel{{}^\bullet\! \Delta}
\def\deld{\Delta^{\hskip -.5mm \bullet}}
\def\dddel{{}^{\bullet \bullet} \! \Delta}
\def\ddeld{{}^{\bullet}\! \Delta^{\hskip -.5mm \bullet}}
\def\deldd{\Delta^{\hskip -.5mm \bullet \bullet}}

\def\la{\langle}
\def\ra{\rangle}

\def\t{\tau}
\def\s{\sigma}
\def\r{\rho}

\def\R{\rightarrow}

\begin{document}

\begin{flushright}
\begin{minipage}{0.25\textwidth} hep-th/0010118 \\
YITP-00-63
\end{minipage}
\end{flushright}
\begin{center}
\bigskip\bigskip\bigskip
{\bf\Large{6D trace anomalies from quantum mechanical \\ \vskip .2cm 
path integrals}}
\vskip 1cm
\bigskip
Fiorenzo Bastianelli $^a$\footnote{E-mail: bastianelli@bo.infn.it} and
Olindo Corradini $^b$\footnote{E-mail: olindo@insti.physics.sunysb.edu} 
\\[.4cm]
{\em $^a$ Dipartimento  di Fisica, Universit\`a di Bologna \\ 
and  INFN, Sezione di Bologna\\ 
via Irnerio 46, I-40126 Bologna, Italy} \\[.4cm]
{\em $^b$ C. N. Yang Institute for Theoretical Physics \\
State University of New York at Stony Brook \\
Stony Brook, New York, 11794-3840, USA}\\
\end{center}
\baselineskip=18pt
\vskip 2.3cm

\centerline{\large{\bf Abstract}}
\vspace{.4cm}

We use the recently developed dimensional regularization (DR) scheme for 
quantum mechanical path integrals in curved space and with a finite time 
interval to compute the trace anomalies for a scalar field in six dimensions.
This application provides a further test of the DR method applied to 
quantum mechanics. It shows the efficiency in higher loop computations of 
having to deal with covariant counterterms only, as required by the DR scheme.

\newpage


\section{Introduction}

Quantum mechanical (QM) path integrals have been usefully applied 
to the computation of chiral 
\cite{Alvarez-Gaume:1983at,Alvarez-Gaume:1984ig}
and trace
\cite{Bastianelli:1992be,Bastianelli:1993ct}
anomalies. 
In these applications the anomalies are identified as certain 
QFT path integral jacobians 
\cite{Fujikawa:1980vr},
first reinterpreted as quantum mechanical 
traces and then given a path integral representation.

The topological character of chiral anomalies explains the relative 
easiness of computing the corresponding QM path integrals: 
the interpretation of chiral anomalies
as indices of certain differential operators
shows how the leading semiclassical approximation of the corresponding 
QM path integrals will give directly the desired result.
On the contrary, the calculation of trace anomalies requires to control 
the full perturbative expansion of QM path integrals on curved spaces.
The latter has been a favorite topic of study over the years.
Most of the early literature dealt with ways of deriving discretized 
expressions, but the program of taking the continuum 
limit till the end to identify the correct regularization scheme to be used 
directly in the continuum was almost never 
completed\footnote{One exception is the description of 
the phase-space path integral in the book of Sakita 
\cite{Sak}: 
it contains noncovariant 
counterterms and the Feynman rules given there can be used to compute to 
any desired loop order since no ambiguous product of distributions is ever 
to be found in the loop expansion 
\cite{Bastianelli:1998jm}. 
The same cannot be said for 
the configuration space version: ambiguities are present there and 
must be resolved with a consistent scheme for multiplying distributions, 
which is what a regularization scheme provides.}.

Thus, starting from refs. 
\cite{Bastianelli:1992be,Bastianelli:1993ct}
a critical re-examination of the correct 
definitions of QM path integrals was initiated which lead to two 
well-defined and consistent schemes of regulating and computing:
mode regularization (MR) 
\cite{Bastianelli:1992be,Bastianelli:1993ct,Bastianelli:1998jm}
and time slicing (TS) 
\cite{deBoer:1995hv,Bastianelli:1998jm}.
Recently, following the suggestion in ref. 
\cite{Kleinert:1999aq}
of using dimensional 
regularization a third way of properly defining the path integrals 
has been developed in 
\cite{Bastianelli:2000nm}: 
the dimensional regularization scheme (DR).
While the counterterms required in MR and TS are noncovariant,
they happen to be covariant in the DR scheme
\cite{Bastianelli:2000nm,Bastianelli:2000pt,Kleinert:1999aq}.

It is the purpose of this paper to test further the consistency of this 
new scheme and show the technical advantage of having to deal with a
manifestly covariant action in performing higher loop calculations 
for 0+1 nonlinear sigma models 
on a finite time interval (i.e. quantum mechanics on curved spaces with 
a finite propagation time). We apply the DR regulated path integral 
to compute the trace anomaly of a conformal scalar in six dimensions.
The correct full trace anomaly for such a scalar (and also other
six dimensional conformal free fields) has only recently been
calculated in 
\cite{Bastianelli:2000hi}
by using the heat kernel results of Gilkey \cite{Gil}.
With a DR path integral calculation we are going to reproduce the 
complete expression of this anomaly.

The paper is structured as follows.
In section 2 we review the DR scheme, in section 3 we use it to 
calculate the trace anomaly for a conformal scalar field in 6D and 
in section 4 we present our conclusions.
Finally, in appendix A we report a list of structures and integrals 
employed in the main text: since the complete calculation is 
somewhat lengthy, it is useful for comparison purposes and future 
reference to record intermediate results.

\section{Dimensional regularization of the path integral}

First, we briefly review the quantization with path integrals of the motion 
of a (non relativistic and unit mass) particle on a curved space with metric 
$g_{ij}(x)$ and scalar potential $V(x)$.
The model is described by the euclidean action\footnote{We perform a 
Wick rotation on the time variable and work 
consistently in the euclidean framework. We also set $\hbar = 1$.}
\bea
S[x^i] = \int_{t_i}^{t_f} \!\!\! dt\ 
\biggl [ {1\over 2} g_{ij}(x)\dot x^i \dot x^j +  V(x) \biggr ].
\label{model}
\ena
In canonical quantization one must choose an ordering consistent with
reparametrization invariance to produce a quantum hamiltonian
$H=-{1\over 2} \nabla^2 +\alpha R + V$ (our curvature conventions 
are found in appendix A).
The value of the parameter $\alpha$ depends on the particular order
chosen 
\cite{DeWitt:1957at}
and conventionally can be taken to vanish with the
agreement of reintroducing the coupling to $R$ through the potential $V$.

Using path integrals the canonical ordering ambiguities re-emerge as the 
need of specifying a regularization scheme. The 1D sigma model 
in eq.~(\ref{model}) contains double derivative interactions
which make Feynman graphs superficially divergent at one and two
loops. However, the nontrivial path integral measure can be exponentiated 
using ghost fields: their effect is to make finite the sum of the Feynman 
graphs but a regularization scheme is still necessary to render finite
each individual divergent graph.
Different regularization schemes require different counterterms to reproduce 
a quantum hamiltonian with $\alpha =0$. In mode regularization and 
time slicing such counterterms are noncovariant. 
In the DR scheme the counterterm is covariant and equal 
to $V_{DR}= {R\over 8}$, as demonstrated in 
\cite{Bastianelli:2000nm,Bastianelli:2000pt}.

Now, let us describe the DR scheme which we are going to apply in the 
next section.
First of all, we find it convenient to use a rescaled time parameter
$\tau$ by defining $t = \beta \tau + t_f$
and $\beta= t_f- t_i$, so that $\tau$ will
take values on the finite interval $I \equiv [-1,0]$.
Then, we introduce bosonic $a^i$ and fermionic $b^i,c^i$ ghosts
to exponentiate the nontrivial part of the 
path integral measure: integrating them
back will formally reproduce the $\prod \sqrt{{\rm det}g_{ij}}$
factor of the measure.
Finally, we introduce $D$ extra infinite regulating dimensions
${\bf t}= (t^1,\ldots,t^D)$ with the prescription that one will take
the limit $D\R 0$ at the very end of all calculations. 
Denoting  $t^\mu \equiv (\t, {\bf t})$ with $\mu=0,1,\ldots,D$ and 
$d^{D+1}t = d\t d^{D}{\bf t }$,
the action in $D+1$ dimensions reads
\bea
S[x,a,b,c]={1\over \beta} \int_\Omega d^{D+1}t 
\left[{1\over 2}\, g_{ij}(x)\left(
\partial_\mu x^i \partial_\mu x^j 
+a^i a^j +b^i c^j\right)+ \beta^2 V(x)+ \beta^2 V_{DR}(x)
\right]
\label{eq:action}
\eea
where $V_{DR}={R\over 8}$ is the counterterm in dimensional regularization
and $ \Omega =I \times R^D$ is the region of
integration containing the finite interval $I$.

The perturbative expansion can be generated by first
decomposing the paths $x^i(\tau)$ into a classical part $x^i_{cl}(\tau)$
satisfying the boundary conditions and quantum fluctuations $q^i(\tau)$
which vanish at the boundary (the ghost fields are 
taken to vanish at the boundary as well) and then
decomposing the lagrangian into a free part plus interactions.
The latter step is achieved by Taylor expanding the metric and the potential
around a fixed point, which we choose to be the final point $x_f$.
Thus, the propagators are recognized to be
\bea
\la x^i(t) x^j(s)\ra &=&
-\beta\ g^{ij}(x_f)
\ \del(t,s)
\\
\la a^i(t) a^j(s)\ra &=&  \beta\ g^{ij}(x_f)\
\Delta_{gh}(t,s) , \ \ \ \ \ 
\la b^i(t) c^j(s)\ra = -2\beta\ g^{ij}(x_f)\
\Delta_{gh}(t,s)
\nonumber
\eea
with
\bea \hskip -.3cm
\del(t,s)&=&\int {d^D{\bf k}\over (2\pi)^D} \sum_{n=1}^\infty 
{-2\over (\pi n)^2+{\bf k}^2}{\rm sin}(\pi n\tau) {\rm sin}(\pi n\sigma)
{\rm e}^{i{\bf k}\cdot ({\bf t}-{\bf s})}  \\
\hskip -.3cm
\del_{gh}(t,s)&=&\int {d^D{\bf k}\over (2\pi)^D} \sum_{n=1}^\infty 
2\, {\rm sin}(\pi n\tau) {\rm sin}(\pi n\sigma)
{\rm e}^{i{\bf k}\cdot ({\bf t}-{\bf s})} =
\delta^{D+1}(t,s) =\delta (\t, \s) \delta^D ({\bf t} -{\bf s}) 
\ena
where 
\bea
\delta(\tau,\sigma) = 
\sum_{n=1}^{\infty} 2\, {\rm \sin}(\pi n\tau){\rm \sin}(\pi n\sigma)
\eea
is the Dirac delta on the space of functions vanishing at $\t ,\s=-1,0 $.
Of course, the function $\del(t,s)$ satisfies the Green equation
\bea
\partial_\mu^2\del(t,s) = \del_{gh}(t,s)=
\delta^{D+1}(s,t) .
\label{green}
\eea
The $D \R 0$ limits of these propagators are the usual ones
\bea
\del(\tau,\sigma) &=& \tau (\sigma + 1 ) \theta (\tau-\sigma)
+ \sigma (\tau + 1) \theta (\sigma-\tau) \\
 \del_{gh}(\t,\s) &=& \dddel(\tau,\sigma)=\delta(\tau,\sigma) 
\eea
where dots on the left/right side denote derivatives with respect
to the first/second variable, respectively.
However, such limits can be used only after one has cast the integrands 
corresponding to the various Feynman diagrams in an unambiguous from by 
making use of the manipulations allowed by the regularization scheme.
In particular, in DR one can use partial integration: it is 
always allowed in the extra $D$ dimension 
because of momentum conservation, while it can be 
performed along the finite time interval whenever there is an explicit 
function which vanishes at the boundary (for example the propagator
of the coordinates $\del(t,s)$).
Along the way one may find terms of the form 
$\partial_\mu^2\del(t,s) $ which according to eq. (\ref{green})
gives Dirac delta functions. The latter can be safely used at the regulated 
level, i.e. in $D+1$ dimensions.
By performing such partial integrations one tries to arrive at an 
unambiguous form of the integrals which can be safely and easily 
calculated even after the limit $D\R 0$ is taken.
 
An explicit example will suffice to describe how the above rules 
are concretely used:
\bea
&&  
\int_{-1}^{0}  \!\!\! d\tau \! \int_{-1}^{0}  \!\!\!  d\sigma \ \ 
(\ddel) ~ (\deld) ~ ( \ddeld ) 
\rightarrow 
\int d^{D+1}t \int d^{D+1}s ~  
( _\mu{\del}) ~  (\del_\nu) ~ (_\mu{\del_\nu})
\cr && 
= 
\int d^{D+1}t \int d^{D+1}s ~  
( _\mu{\del}) ~  _\mu\biggl({1\over 2}(\del_\nu)^2 \biggr )
= - {1\over 2}  \int d^{D+1}t \int d^{D+1}s ~ 
( _{\mu\mu}{\del}) ~ (\del_\nu)^2  
\cr && 
= - {1\over 2} 
\int d^{D+1}t \int d^{D+1}s ~ 
\delta^{D+1}(t,s) ~ (\del_\nu)^2  
= - {1\over 2}  \int d^{D+1}t ~ (\del_\nu)^2|_t 
\cr && 
\rightarrow  - {1\over 2} 
\int_{-1}^{0}  \!\!\! d\tau \ (\deld)^2|_\tau 
= -{1\over 24} \nonumber 
\ena 
where the symbol $ |_\t$ means that one should set $\s =\t$.

Thus, we see that the rules of computing in DR are quite similar to those 
used in MR, the only diversity being in the different options allowed in
partial integrations.
In DR the rule for contracting which indices with which indices 
follows directly from the regulated
action in (\ref{eq:action}) and only certain partial 
integrations are allowed in $D+1$ dimensions.
In MR one regulates by cutting off all mode sums at a
large mode $N$ and then performs partial integrations: 
all derivatives are now of the same nature and different options
of partial integrations arise.
This explains the origin of different counterterms for these two 
regularizations.

\section{Trace anomalies for a conformal scalar in 6D}
As described in 
\cite{Bastianelli:1992be,Bastianelli:1993ct},
one-loop trace anomalies can be obtained by computing a certain 
Fujikawa jacobian suitably regulated and represented as a quantum 
mechanical path integral with periodic boundary conditions
\bea
\int d^6x\ \sqrt{g}\ \sigma(x) \langle T^a{}_a (x) \rangle=
\lim_{\beta \R 0}
{\rm Tr} [ \sigma\ {\rm e}^{-\beta H}] = \lim_{\beta \R 0}
\int_{_{PBC}} \hskip -.6cm {\cal D}x\ \s(x)\ {\rm e}^{-S[x]} ,
\label{pbc}
\eea
where on the left hand side $T^a{}_a$ denotes the trace of the 
stress tensor for a 6D conformal scalar and $\sigma(x)$
is an arbitrary function describing an infinitesimal Weyl variation.
In the first equality the infinitesimal part of the Fujikawa 
jacobian has been regulated with the conformal scalar field
kinetic operator $H = -{1\over 2}\nabla^2 -{1\over 10}R$.
The limit $\beta \R 0$ should be taken after removing divergent terms 
in $\beta$ (which is what the renormalization of the scalar field 
QFT will do), and so it picks up just the $\beta$ independent term.
Finally on the right hand side the trace is given a representation as
a path integral corresponding to a model with hamiltonian $H$ and
with periodic boundary conditions.
The latter can be obtained using the quantum mechanics
described in the previous section 
with a scalar potential $V= -{1\over 10}R$.

Thus we start computing the terms in the loop expansion
of the path integral described in section 2.
It will soon be clear that it is enough to compute up to order
$\beta^3$ i.e. up to 4 loops ($\beta$ can be taken as the loop counting 
parameter, as evident form eq. (\ref{eq:action})).
We use reparametrization invariance and choose
Riemann normal coordinates centered at the point $x^i_0$
representing the boundary conditions at $\tau=-1,0$,
and which will be integrated over to recreate the full
periodic boundary conditions on the right hand side of (\ref{pbc}).

The expansion of the metric in Riemann normal coordinates is well-known.
For our case, since the action including the counterterm
is manifestly covariant, that expansion can be easily generated by the 
method described for this context in 
\cite{Bastianelli:1992be}.
One obtains the following terms needed in our approximation
\bea
 g_{mn}(x) dx^m dx^n &=& 
\biggl[
g_{mn} + {1\over 3}  R_{mabn} x^a x^b 
+ {1\over 3!} \nabla_i R_{mabn} x^a x^b x^i
\nonumber \\ && 
+ {6\over 5!}\biggl( \nabla_i \nabla_j  R_{mabn} + {8\over 9}
 R_{mab}{}^{\alpha}  R_{\alpha ijn}\biggr) x^a x^b x^i x^j 
\nonumber \\ &&
+ {1\over 5!}{4\over 3} 
\biggl( \nabla_i \nabla_j  \nabla_k R_{mabn} 
+ 4 R_{mab}{}^{\alpha} \nabla_i    
R_{\alpha jkn}\biggr)
 x^a x^b x^i x^j x^k \nonumber \\ &&
+ {10\over 7!}\biggl(  \nabla_i \nabla_j  \nabla_k \nabla_l R_{mabn}
+ {34\over 5}  R_{m ij}{}^{\alpha} \nabla_k \nabla_l R_{\alpha ab n}  
+ {11\over 2} \nabla_i R_{m ab}{}^{\alpha} \nabla_j R_{\alpha kl n}
\nonumber \\ &&
+ {8\over 5}  R_{m ab}{}^{ \alpha} R_{\alpha ij}{}^{\beta}
R_{\beta kl n}\biggr)  x^a x^b x^i x^j x^k x^l +\cdots
\biggr]dx^m dx^n \nonumber \\ 
V(x) &=& V+ (\nabla_i V) x^i +{1\over 2}(\nabla_i\nabla_j V) x^ix^j 
+{1\over 3!} (\nabla_i\nabla_j\nabla_k V) x^i x^j x^k \nonumber \\ &&
+{1\over 4!} (\nabla_i\nabla_j\nabla_k\nabla_l V) x^i x^j x^k x^l
+\cdots
\eea
where all tensorial quantities on the right hand sides 
are evaluated at the origin of the coordinate system. Notice that
for the MR and TS regularization schemes the counterterms
are noncovariant and their expansions cannot be generated so easily:
obtaining the vertices from those counterterms
would require a tedious computation.

Plugging the above expansions in the action (\ref{eq:action})
and noticing that the factor $\beta^2$ raises by two the loop order 
for each vertex coming from the potential or the counterterm,
we compute~\footnote{Since $x^i(-1)=x^i(0)\equiv x_0$ the classical field is
$x_{cl}(\t)=0$; hence all diagrams with external fields vanish.
We indicate with $S_n$ the interaction terms containing 
$n$ fields when originating from the expansion of the 
metric and $n-4$ fields when originating from the scalar potential.} 
\bea
&& \langle x_0,\beta | x_0,0 \rangle  =
\int  {\cal D} x
\ e^{ - S }
= A\ 
\langle {\rm e}^{- S_{int}} \rangle \nonumber\\
&& = A\ \exp \biggl [ 
- \langle S_4 \rangle
- \langle S_6 \rangle
- \langle S_8 \rangle
+ {1\over 2}  \langle S_4^2 \rangle_c
+ {1\over 2}  \langle S_5^2  \rangle_c
+ \langle S_4 S_6 \rangle_c
- {1\over  6}  \langle S_4^3  \rangle_c
+ O(\beta^{4}) \biggr ]
\label{espa}
\eea
where the subscript ``c'' stands for connected diagrams only and where 
$A = (2 \pi \beta)^{-{D\over 2}}$ gives the correct normalization 
of the path integral
measure. Because of this normalization we see 
that for $D=6$ the $\beta$-independent term is obtained by picking
up the $\beta^3$ contributions from the expansion of the exponential
on the right hand side of eq. (\ref{espa}).

The terms up to 3 loops are easily computed in DR
by using the detailed expressions reported in 
\cite{Bastianelli:1999jb}: one just needs to compute the integrals reported 
there using the DR rules. Including for simplicity the counterterm inside 
the potential $V$, we obtain
\bea 
&& \langle S_4 \rangle = -\beta \biggl[{1\over 24} R - V\biggr] \\
&& \langle S_6 \rangle =  - {\beta^2\over 12}
\biggl[{1\over 40} \nabla^2 R +{1\over 90} R_{mn}^2 +{1\over 60} R^2_{mnab}
 - \nabla^2 V \biggr] \\
&& \langle S_4^2 \rangle_c = - {\beta^2\over 72} \biggl[{1\over 3} R_{mn}^2
\biggr].
\ena
To achieve notational simplicity in the remaining 4-loop terms we use  
the basis of curvature invariants given in appendix A and compute
the terms reported there. We obtain
\bea
\langle S_8 \rangle &=& -{\beta^3 \over 7!} 
\biggl[{17\over 15} K_4 -{16\over 15}K_5 +{8\over 5}K_6 +{5\over 12}K_7 
-{8\over 3}K_8 +{11\over 10} K_{10} +{3\over 2}K_{11}
\nonumber \\ && 
-{19\over 20}K_{12}+{149\over 48}K_{13}
-{25\over 8}K_{14} +{11\over 16} K_{15} -{5\over 24} K_{16}
+{3\over 8}K_{17}
\biggr]
\nonumber \\ && 
-{\beta^3 \over 6!} \biggl[ 2 R^{mn}\nabla_m\nabla_n V +
\nabla^m R \nabla_m V - 3 \nabla^4 V \biggr] \\
\la S_5^2\ra_c &=&- {\beta^3\over 6!}
\biggl[
{23\over 24}K_{13}-{3\over 4}K_{14}-{1\over 8}K_{15}
-{5\over 48}K_{16} +5\nabla^mR\nabla_mV- 60(\nabla_mV)^2
\biggr] \\
\langle S_4 S_6 \rangle_c &=&-{\beta^3 \over 6!}
\biggl[{13\over 45}K_4-{1\over 5}K_5+{2\over 15}K_6
+{3\over 10}K_{10}-{1\over 10}K_{12}\biggr]
\label{eq:S4S6} \\
\langle S_4^3  \rangle_c &=& -{\beta^3\over 6!}
\left[{2\over 3}K_4+{1\over 6}K_7+{4\over 3}K_8
\right].
\label{eq:S4cubed}
\ena

Inserting all these values into eq. (\ref{espa}) we get
\bea
\langle x_0,\beta | x_0,0 \rangle  &=&
{1\over (2\pi \beta)^3}  \exp \biggl [ \beta \biggl({1\over 24} R - V\biggr)
+\beta^2 \biggl({1\over 720} (R^2_{mnab}-R_{mn}^2)+{1\over 480} \nabla^2 R 
 - {1\over 12}\nabla^2 V \biggr)
\nonumber \\
&&+{\beta^3 \over 8!} 
\biggl(-{8\over 9} K_4 +{8\over 3}K_5 +{16\over 3}K_6 +{44\over 9}K_7 
-{80\over 9}K_8 - 8 K_{10} + 12K_{11} - 2 K_{12}
\nonumber \\ && 
-2K_{13}-4K_{14} +9 K_{15} +{5\over 4}K_{16}+3 K_{17} \biggr)
\nonumber \\ && 
+{\beta^3 \over 6!} 
\biggl( 2 R^{mn}\nabla_m\nabla_n V -{3\over 2}
\nabla^m R \nabla_m V 
+ 30 (\nabla_m V)^2 - 3 \nabla^4 V \biggr)
+ O(\beta^{4}) \biggr ].
\eea

Now, using the value $V={1\over 8}R -{1\over 10}R ={1\over 40}R$
to take into account the counterterm $V_{DR}$ and the conformal coupling
of the scalar field, we compare with eq. (\ref{pbc}) and obtain the 
corresponding trace anomaly 
\bea
\langle T^a{}_a \rangle &=& {1\over (2\pi)^3} {1\over 8!}\biggl[
{7\over 225} K_1 
-{14\over 15} K_2 
+{14\over 15} K_3 
-{8\over 9} K_4
+{8\over 3} K_5
+{16\over 3} K_6
+{44\over 9} K_7
-{80\over 9} K_8 \nonumber \\
&&-8 K_{10}
+12 K_{11}
+{4\over 5}K_{12}
-2 K_{13}
-4 K_{14}
+9 K_{15}
+{1\over 5} K_{16}
-{6\over 5} K_{17}
\biggr].
\eea
It agrees with the one given in \cite{Bastianelli:2000hi},
where it was shown that it can be cast also as
\bea
\langle T^a{}_a \rangle &=& {1\over (2\pi)^3} {1\over 8!}\biggl[
-{5\over 72}E_6 -{28\over 3} I_1 +{5\over 3}I_2 + 2I_3 +
{\rm trivial \ anomalies}
\biggr]
\eea
with the topological Euler density given by
\bea
E_{6} &=& -\epsilon_{m_1n_1m_2n_2m_3n_3}\epsilon^{a_1b_1a_2b_2a_3b_3}
R^{m_1n_1}{}_{a_1b_1}R^{m_2n_2}{}_{a_2b_2} R^{m_3n_3}{}_{a_3b_3}
\ena
and the three Weyl invariants
\bea
I_{1} &=& C_{amnb} C^{mijn} C_{i}{}^{ab}{}_j \\
I_{2} &=& C_{ab}{}^{mn} C_{mn}{}^{ij} C_{ij}{}^{ab} \\
I_{3} &=& C_{mabc}\left(\nabla^{2}\delta^{m}_n+4R^{m}_n
-\frac{6}{5}R\delta^{m}_n\right)C^{nabc}+{\rm trivial \ anomalies},
\ena
whereas
the coefficients of the trivial anomalies are unimportant since 
they can be changed by the variation of local counterterms.
The structure of trivial anomalies has been fully analyzed
in \cite{Bastianelli:2000rs}.
It is interesting to note, after inspecting the results in 
\cite{Bastianelli:2000rs}, that the coefficients of
$K_1,\ K_2$ and $ K_3$ never appear in the trivial anomalies. 
At the same time they are produced in the previous calculation 
by disconnected diagrams. Thus one may fix three
of the four true anomalies by a simpler lower loop
calculation, while the remaining independent fourth nontrivial anomaly,
which can be taken as the one corresponding to $E_6$,
could be fixed by an independent calculation on the simplified geometry
of a maximally symmetric space.

\section{Conclusions}

We have used the recently developed dimensional regularization scheme 
for quantum mechanical path integrals \cite{Bastianelli:2000nm}
to compute the trace anomaly for a scalar field in six dimensions.
The identification of the full anomaly required a complete 4-loop 
quantum mechanical computation.
Technically, the covariance of the counterterm $V_{DR}$ 
allows a more efficient identification of the corresponding vertices
than in the MR and TS regularization schemes.

We noticed that the coefficients of three of the four nontrivial anomalies
could as well be obtained by a simpler  3-loop calculation.
One may speculate that such a fact may happen also for $D=8$
trace anomalies: there one would need to compute the quantum mechanics
up to 5-loops, but it could happen that all nontrivial 
anomalies but one could be fixed by a simpler 4-loop calculation
(presented in this paper) and the remaining one by a calculation on a 
simplified geometry. 
To concretely check this conjecture, one would need a cohomological
analysis to identify the structure of all trivial and nontrivial 
anomalies, as the one given in ref. \cite{Bastianelli:2000rs}
for the six dimensional case.
However, such an analysis is not available in the literature yet.

One could couple the nonlinear sigma model
to non-abelian gauge potentials to 
obtain the trace anomalies of other six dimensional conformal fields
\cite{Bastianelli:2000hi}. In such an extension
the main new complication is related to the time ordering 
prescription to be used for achieving gauge covariance, as 
employed in \cite{Bastianelli:1993ct}, 
which forces to compute different DR integrals for different ordering of 
the vertices. An approach which could guarantee non-abelian covariance
in a more straightforward manner would clearly be welcome.
It may be related to the extra ghost fields used in 
\cite{deBoer:1995hv}.

While we have justified our anomaly computation by viewing it as the
calculation of a certain Fujikawa jacobian, conceptually it can be thought of
as performed in the first quantized approach of the scalar
particle theory (see the discussion in  \cite{Bastianelli:1994xq}).
Given that interpretation, it would be interesting to investigate
if such worldline path integrals in curved space could be useful to
simplify computations of scattering amplitudes and effective actions 
of perturbative QFT coupled to gravity, as it happens
in the flat space case \cite{wla}.

\vfill\eject
\section*{Appendix A}

We use the convention $[\nabla_a,\nabla_b]V^c=R_{ab}{}^c{}_dV^d$, 
$R_{ab}=R_{ac}{}^c{}_b$. 
It is useful for notational purposes to introduce a 
basis of curvature invariants cubic in the curvature
\EQ
\begin{array}{lll}
K_1 = R^3  & 
K_2 = R R_{ab}^2 & 
K_3 = R R_{abmn}^2 \\ [3mm] 
K_4 = R_a{}^m R_m{}^i R_i{}^a & 
K_5 =  R_{ab} R_{mn} R^{mabn} & 
K_6= R_{ab} R^{amnl} R^b{}_{mnl} \\ [3mm] 
K_7 = R_{ab}{}^{mn} R_{mn}{}^{ij}R_{ij}{}^{ab} \ \ \ \ &
K_8 = R_{amnb} R^{mijn} R_{i}{}^{ab}{}_j \ \ \ \ &
K_9 = R\nabla^2 R \\ [3mm]
K_{10} = R_{ab}\nabla^2 R^{ab} &
K_{11} = R_{abmn}\nabla^2 R^{abmn} &
K_{12} = R^{ab} \nabla_a \nabla_b R \\ [3mm]
K_{13} = (\nabla_a R_{mn})^2  &
K_{14} = \nabla_a R_{bm} \nabla^b R^{am} &
K_{15} = (\nabla_i R_{abmn})^2  \\ [3mm]
K_{16} = (\nabla_a R)^2  &
K_{17} =\nabla^4 R . & 
\label{inv}
\end{array}
\EN 
It differs from the basis used in \cite{Bastianelli:2000rs,Bonora:1986cq}
only in the definition of $K_{16}$: the one used above enters more 
naturally in our calculations.

In the main text contributions of order $\beta^3$ 
to the effective action come from the terms listed below.
In the list of integrals we use the following conventions.
The limits of integration are $[-1,0]$ for all variables.
For 3-dimensional integrals
the first group of propagator in round brackets
depends on $(\t,\s)$, the second on 
$(\s,\rho)$ and the third on $(\rho,\t)$, with this precise order,
while terms at coinciding points are explicitly indicated.
For 2-dimensional integrals the propagators at non-coinciding points
depend on  $(\t,\s)$, while for 1-dimensional integrals
all terms are obviously taken at coinciding points.
We use the shorthand notation $\overline{\ddeld} = \ddeld+\dddel$.
The DR regularization is immediate and we quote the DR values.
\vskip 1cm

{\small
\noindent $\bullet$\ \underline{$\langle S_4^3\rangle_c$}
\bea
\langle S_4^3  \rangle_c &=& A_0 + A_1 + A_2 + A_3 
\\ 
A_0 &=& {\beta^3 \over 9} \biggl[
\biggl ({1\over 4}K_7 + 2 K_8 \biggr )
\biggl (I_1^{A0} -I_2^{A0} -I_3^{A0} +2 I_4^{A0} +2I_5^{A0} -2I_6^{A0} 
+{1\over 3}I_{11}^{A0}\biggr)
\nonumber \\ && 
- \biggl ({7\over 2}K_7 + K_8 \biggr ) 
\biggl({1\over 3}I_7^{A0} + I_9^{A0} \biggr)
+ \biggl ({13\over 4}K_7 - K_8 \biggr )
\biggl(I_8^{A0} +{1\over 3}I_{10}^{A0}\biggr)
\biggr ]
\\
A_1 &=& {\beta^3 \over 6} \ K_6
(I_1^{A1} - I_2^{A1} - 2 I_3^{A1} + 2 I_4^{A1} + I_5^{A1} - I_6^{A1} 
+ I_7^{A1} - I_8^{A1} - 2 I_9^{A1} + 2 I_{10}^{A1} 
\nonumber \\
&& 
+ I_{11}^{A1} - I_{12}^{A1} 
- 2 I_{13}^{A1} + 2I_{14}^{A1} +2 I_{15}^{A1} - 2 I_{16}^{A1} 
+ 2 I_{17}^{A1} - 2 I_{18}^{A1} - 2 I_{19}^{A1} + 2 I_{20}^{A1} )
\\
A_2 &=& {\beta^3 \over 9}  K_5 
[I_1^{A2} - I_2^{A2} + I_3^{A2} - I_4^{A2} + 4 I_5^{A2} - I_6^{A2}
 - 2 I_7^{A2} -I_8^{A2} 
+ I_9^{A2} 
\nonumber \\
&&+ I_{10}^{A2} +2  (- I_{11}^{A2} - I_{12}^{A2} - I_{13}^{A2} + I_{14}^{A2}
+I_{15}^{A2} -I_{16}^{A2}-I_{17}^{A2} + I_{18}^{A2} + I_{19}^{A2})]
\\
A_3 &=& {\beta^3 \over 27}K_4
(I_1^{A3}-6I_2^{A3}+3I_3^{A3}+3I_4^{A3}-6I_5^{A3}-6I_6^{A3}+3I_7^{A3}+3I_8^{A3}
+6I_9^{A3}\nonumber\\
&&+I_{10}^{A3}-6I_{11}^{A3}+3I_{12}^{A3}+3I_{13}^{A3}+6I_{14}^{A3}-2I_{15}^{A3}
-6I_{16}^{A3})
\ena

Integrals in $A_0$:
\EQ
\begin{array}{ll}
\hskip-.5cm
I^{A_0}_{1} = \3i (\deld{}^2)(\del^2)(\ddeld{}^2-\dddel{}^2)=-{13\over 3780} 
& \hskip -2.5cm
I^{A_0}_{2} = \3i (\del\ \deld)(\ddel\ \del)(\ddeld{}^2-\dddel{}^2)
=-{1\over 15120} \\ [2mm]\hskip-.5cm
I^{A_0}_{3} = \3i (\ddeld\ \deld)(\del^2)(\ddeld\ \ddel)
={13\over 7560} &\hskip -2.5cm
I^{A_0}_{4} = \3i (\ddeld\ \del)(\ddel\ \del )(\ddeld\ \ddel)
=-{13\over 15120}  \\ [2mm]\hskip-.5cm
I^{A_0}_{5} = \3i (\ddel\ \deld)(\ddel\ \del)(\ddeld\ \ddel)
=-{1\over 3780} &\hskip -2.5cm
I^{A_0}_{6} = \3i (\ddel\ \del)(\ddel{}^2)(\ddeld\ \ddel)
={1\over 1890}  \\ [2mm]\hskip-.5cm
I^{A_0}_{7} = \3i [(\del\ \ddeld)(\del\ \ddeld)(\del\ \ddeld) +
 (\del\ \dddel)(\del\ \dddel)(\del\ \dddel)]
=-{43\over 15120}  & \\[2mm]\hskip-.5cm
I^{A_0}_{8} = \3i (\ddel\ \deld)(\del\ \ddeld)(\del\ \ddeld)
={11\over 15120}  &\hskip -2.5cm
I^{A_0}_{9} = \3i (\ddel\ \deld)(\ddel\ \deld)(\del\ \ddeld)
={1\over 756}  \\[2mm]\hskip-.5cm
I^{A_0}_{10} = \3i (\ddel\ \deld)(\ddel\ \deld)(\ddel\ \deld)
=-{1\over 945} &\hskip -2.5cm
I^{A_0}_{11} = \3i (\ddel^2)(\ddel^2)(\ddel^2)={1\over 756}  
\end{array}
\EN 

Integrals in $A_1$:
\EQ
\begin{array}{ll}
\hskip-.5cm
I^{A_1}_1 =\3i (\overline{\ddeld})|_\t (\deld)(\ddeld\ \del^2)(\ddel)
=-{13\over 3780} &\hskip-1.5cm
I^{A_1}_2 =\3i (\overline{\ddeld})|_\t (\deld)(\ddel\ \del\ \deld)
(\ddel)={1\over 1890}\\[2mm]
\hskip-.5cm
I^{A_1}_3 =\3i (\overline{\ddeld})|_\t (\deld)(\ddeld\ \del\ \deld)(\del)
=-{1\over 945} &\hskip-1.5cm
I^{A_1}_4 =\3i (\overline{\ddeld})|_\t (\deld)(\ddel\ \deld{}^2)(\del)
=-{1\over 3780}\\[2mm]\hskip-.5cm
I^{A_1}_5 =\3i (\overline{\ddeld})|_\t (\del)(\del[\ddeld{}^2-\dddel^2])
(\del)={1\over 432} &\hskip-1.5cm
I^{A_1}_6 =\3i (\overline{\ddeld})|_\t (\del)(\ddeld\ \deld\ \ddel)
(\del)=-{1\over 15120}\\[2mm]\hskip-.5cm
I^{A_1}_7 =\3i (\del)|_\t [(\ddeld)(\del^2\ \ddeld)(\ddeld) +
(\dddel)(\del^2\ \dddel)(\dddel)]=-{1\over 216}&\\[2mm]
\hskip-.5cm
I^{A_1}_8 =\3i (\del)|_\t (\ddeld)(\ddel\ \del\ \deld)(\ddeld)
=-{1\over 15120}&\\[2mm]\hskip-.5cm
I^{A_1}_9 =\3i (\del)|_\t (\ddeld)(\ddeld\ \del\ \deld)(\deld)
=-{1\over 2160}&\hskip-1.5cm
I^{A_1}_{10} =\3i (\del)|_\t (\ddeld)(\ddel\ \deld{}^2)(\deld)
={2\over 945}\\[2mm]\hskip-.5cm
I^{A_1}_{11} =\3i (\del)|_\t (\ddel)(\del [\ddeld{}^2- \dddel^2])(\deld)
=-{19\over 15120}&\hskip-1.5cm
I^{A_1}_{12} =\3i (\del)|_\t (\ddel)(\ddeld\ \ddel\ \deld)(\deld)
=-{1\over 1512}\\[2mm]\hskip-.5cm
I^{A_1}_{13} =\3i (\ddel)|_\t (\ddeld)(\ddeld\ \del^2)(\ddel)
={13\over 7560}&\hskip-1.5cm
I^{A_1}_{14} =\3i (\ddel)|_\t (\ddeld)(\ddel\ \del\ \deld)(\ddel)
=-{1\over 3780}\\[2mm]\hskip-.5cm
I^{A_1}_{15} =\3i (\ddel)|_\t (\ddeld)(\ddeld\ \deld\ \del)(\del)
={17\over 15120}&\hskip-1.5cm
I^{A_1}_{16} =\3i (\ddel)|_\t (\ddeld)(\ddel\ \deld{}^2)(\del)
={1\over 7560}\\[2mm]\hskip-.5cm
I^{A_1}_{17} =\3i (\ddel)|_\t (\ddel)(\ddeld\ \ddel\ \del)(\ddel)
=-{1\over 15120}&\hskip-1.5cm 
I^{A_1}_{18} =\3i (\ddel)|_\t (\ddel)(\ddel^2\  \deld)(\ddel)
={1\over 7560}\\[2mm]\hskip-.5cm
I^{A_1}_{19} =\3i (\ddel)|_\t (\ddel)(\del[\ddeld{}^2- \dddel^2])(\del)
=-{1\over 864}&\hskip-1.5cm
I^{A_1}_{20} =\3i (\ddel)|_\t (\ddel)(\ddeld\ \ddel\ \deld)(\del)
={1\over 30240} 
\end{array}
\EN 

Integrals in $A_2$:
\EQ
\begin{array}{ll}
\hskip-.5cm
I^{A_2}_1 =\3i (\overline{\ddeld})|_\t (\overline{\ddeld})|_\s 
(1)(\del^2)(\ddel^2) ={1\over 756} & \hskip .3cm
I^{A_2}_2 =\3i (\overline{\ddeld})|_\t (\overline{\ddeld})|_\s 
(1)(\del\ \deld)(\ddel\ \del)={1\over 1890}\\[2mm]
\hskip-.5cm
I^{A_2}_3 =\3i (\del)|_\t (\del)|_\s (1)(\ddel^2)(\ddeld{}^2 
- \dddel^2) = -{11\over 1890} & \hskip .3cm
I^{A_2}_4 =\3i (\del)|_\t (\del)|_\s (1)(\ddel\ \ddeld)(\ddeld\ \deld)
={2\over945}\\[2mm]
\hskip-.5cm
I^{A_2}_5 =\3i (\deld)|_\t (\deld)|_\s (1)(\ddel\ \del)(\ddeld\  \ddel) 
={1\over 3024} &\hskip .3cm
I^{A_2}_6 =\3i (\deld)|_\t (\deld)|_\s (1)(\ddel\ \deld)(\ddel\ \deld)
={1\over 30240}\\[2mm]
\hskip-.5cm
I^{A_2}_7 =\3i (\deld)|_\t (\deld)|_\s (1)(\del\ \ddeld)(\ddel\  
\deld) =-{1\over 6048}&\hskip .3cm
I^{A_2}_8 =\3i (\deld)|_\t (\deld)|_\s (1)(\del\ \ddeld)(\del\ 
\ddeld)={5\over 6048}\\[2mm]
\hskip-.5cm
I^{A_2}_9 =\3i (\overline{\ddeld})|_\t (\del)|_\s (1)(\ddel^2)
(\ddel^2)=-{1\over 378}&\hskip .3cm
I^{A_2}_{10} =\3i (\overline{\ddeld})|_\t (\del)|_\s 
(1)(\ddeld{}^2 - \dddel^2)(\del^2)={11\over 3780}\\[2mm]
\hskip-.5cm
I^{A_2}_{11} =\3i (\overline{\ddeld})|_\t (\del)|_\s 
(1)(\ddel\ \ddeld)(\ddel\ \del)=-{1\over 945}&\hskip .3cm
I^{A_2}_{12} =\3i (\overline{\ddeld})|_\t (\deld)|_\s 
(1)(\ddel \del)(\ddel^2)= -{1\over 1512}\\[2mm]
\hskip-.5cm
I^{A_2}_{13} =\3i (\overline{\ddeld})|_\t (\deld)|_\s 
(1)(\ddeld\ \deld)(\del^2)=-{1\over 1512}&\hskip .3cm
I^{A_2}_{14} =\3i (\overline{\ddeld})|_\t (\deld)|_\s 
(1)(\del\ \ddeld)(\ddel\ \del)=-{1\over 1512}\\[2mm]
\hskip-.5cm
I^{A_2}_{15} =\3i (\overline{\ddeld})|_\t (\deld)|_\s 
(1)(\ddel\ \deld)(\ddel\ \del) = {1\over 7560}&\hskip .3cm
I^{A_2}_{16} =\3i (\del)|_\t (\deld)|_\s 
(1)(\ddel \del)(\ddeld{}^2-\dddel^2)=-{11\over7560}\\[2mm]
\hskip-.5cm
I^{A_2}_{17} =\3i (\del)|_\t (\deld)|_\s 
(1)(\ddeld\ \deld)(\deld{}^2) = {1\over 756}&\hskip .3cm
I^{A_2}_{18} =\3i (\del)|_\t (\deld)|_\s 
(1)(\del\ \ddeld)(\ddeld\ \deld)={1\over756}\\[2mm]
\hskip-.5cm
I^{A_2}_{19} =\3i (\del)|_\t (\deld)|_\s 
(1)(\ddel\ \deld)(\ddeld\ \deld)=-{1\over3780}&\hskip .3cm
\end{array}
\EN 

Integrals in $A_3$:
\EQ
\begin{array}{ll}
\hskip-1cm
I^{A_3}_1 =\3i (\overline{\ddeld})|_\t (\overline{\ddeld})|_\s 
(\overline{\ddeld})|_\r
(\del)(\del)(\del) = -{1\over 945}
&\hskip -2.4cm
I^{A_3}_2 =\3i (\overline{\ddeld})|_\t (\overline{\ddeld})|_\s (\ddel)|_\r 
(\del)(\deld)(\del) = {1\over 1890}
\\[2mm]
\hskip-1cm
I^{A_3}_3 =\3i (\overline{\ddeld})|_\t (\overline{\ddeld})|_\s (\del)|_\r 
(\del)(\deld)(\ddel) = {1\over 756}
&\hskip-2.4cm
I^{A_3}_4 =\3i (\overline{\ddeld})|_\t (\del)|_\s (\del)|_\r 
(\deld)(\ddeld)(\ddel) = -{29\over 7560}
\\[2mm]
\hskip-1cm
I^{A_3}_5 =\3i (\overline{\ddeld})|_\t (\deld)|_\s (\del)|_\r 
(\del)(\ddeld)(\ddel) = -{13\over 15120}
&\hskip-2.4cm
I^{A_3}_6 =\3i (\overline{\ddeld})|_\t (\deld)|_\s (\del)|_\r 
(\deld)(\deld)(\ddel)= -{1\over 2160}
\\[2mm]
\hskip-1cm
I^{A_3}_7 =\3i (\overline{\ddeld})|_\t (\deld)|_\s (\deld)|_\r 
(\deld)(\del)(\ddel)= -{11\over 30240}
&\hskip-2.4cm
I^{A_3}_8 =\3i (\overline{\ddeld})|_\t (\deld)|_\s (\deld)|_\r 
(\del)(\ddeld)(\del) = -{11\over 30240}
\\[2mm]
\hskip-1cm
I^{A_3}_9 =\3i (\overline{\ddeld})|_\t (\deld)|_\s (\deld)|_\r 
(\deld)(\deld)(\del)= -{1\over 6048}
&\\[2mm]\hskip-1cm
I^{A_3}_{10} =\3i (\del)|_\t (\del)|_\s (\del)|_\r 
[(\ddeld)(\ddeld)(\ddeld)+(\dddel)(\dddel)(\dddel)]
= -{71\over 7560}
\\[2mm]
\hskip-1cm
I^{A_3}_{11} =\3i (\del)|_\t (\ddel)|_\s (\del)|_\r 
(\ddeld)(\deld)(\ddeld)= {29\over 15120}
&\hskip-2.4cm
I^{A_3}_{12} =\3i (\del)|_\t (\ddel)|_\s (\ddel)|_\r 
(\ddel)(\ddeld)(\deld)= {19\over 30240}
\\[2mm]
\hskip-1cm
I^{A_3}_{13} =\3i (\del)|_\t (\ddel)|_\s (\ddel)|_\r 
(\ddeld)(\del)(\ddeld)= {31\over 30240}
&\hskip-2.4cm
I^{A_3}_{14} =\3i (\del)|_\t (\ddel)|_\s (\ddel)|_\r 
(\ddel)(\ddel)(\ddeld)= -{1\over 6048}
\\[2mm]
\hskip-1cm
I^{A_3}_{15} =\3i (\ddel)|_\t (\ddel)|_\s (\ddel)|_\r 
(\ddel)(\ddel)(\ddel)= -{1\over 60480}
&\hskip-2.4cm
I^{A_3}_{16} =\3i (\ddel)|_\t (\ddel)|_\s (\ddel)|_\r 
(\del)(\ddel)(\ddeld) = {11\over 60480}
\end{array}
\EN 

\vskip 1cm

\noindent $\bullet$\ \underline{$\langle S_5^2\ra_c $}
\bea
\la S_5^2\ra_c &=&-{\beta^3\over 144} \biggl[
K_{15}(6I_1^{B_0}-12I_2^{B_0}+6I_3^{B_0})
+K_{16}(4I_1^{B_2}-8I_2^{B_2}+4I_3^{B_2})+\nonumber\\&&
\hskip-0.5cm+K_{13}(2I_1^{B_1}-8I_2^{B_1}+4I_3^{B_1}+22I_4^{B_1}
-20I_5^{B_1}
-48I_6^{B_1}+40I_7^{B_1}+24I_8^{B_1}-16I_9^{B_1}) \nonumber\\
&&\hskip-0.5cm+K_{14}(4I_1^{B_1}-16I_2^{B_1}+8I_3^{B_1}-20I_4^{B_1}
+24I_5^{B_1}+32I_6^{B_1}-48I_7^{B_1}-16I_8^{B_1}+32I_9^{B_1})\biggr]
\nonumber\\&& 
\hskip-0.5cm-{\beta^3\over 3}\biggl[
3(\nabla_aV)^2I_1^{B_3}+\nabla_kR\nabla^kV
(I_2^{B_3}-I_3^{B_3})\biggr]
\eea

Integrals:
\EQ
\begin{array}{ll}
\hskip-1cm
I_1^{B_0}=\2i\del^3 (\ddeld{}^2-\dddel^2)=0&\hskip1cm
I_2^{B_0}=\2i\del^2\ \ddel\ \ddeld\ \deld={1\over 840}\\[2mm]
\hskip-1cm
I_3^{B_0}=\2i\del(\ddel)^2(\deld)^2=-{1\over 560} &
\hskip1cm I_1^{B_1}=\2i\del^3=-{1\over 560} \\[2mm]\hskip-1cm 
I_2^{B_1}=\2i\ddel|_\s\del^2\deld={1\over 1680}&\hskip1cm
I_3^{B_1}=\2i\del|_\s\ \del\ (\deld)^2={1\over 336} \\[2mm]\hskip-1cm
I_4^{B_1}=\2i\del|_\t\ \del|_\s\ \del(\ddeld{}^2-\dddel^2)
={17\over 2520}&\hskip1cm
I_5^{B_1}=\2i\del|_\t\ \del|_\s\ 
\ddel\ \ddeld\ \deld=-{1\over 2520} \\[2mm]\hskip-1cm
I_6^{B_1}=\2i\del|_\t\ \ddel|_\s\ \ddel\ \ddeld\ \del=-{1\over 720}&\hskip1cm 
I_7^{B_1}=\2i\del|_\t\ \ddel|_\s\ (\ddel)^2\deld=-{1\over 5040}\\[2mm]
\hskip-1cm
I_8^{B_1}=\2i\ddel|_\t\ \ddel|_\s\ \del^2\ \ddeld=-{1\over 1008} &\hskip1cm
I_9^{B_1}=\2i\ddel|_\t\ \ddel|_\s\ \ddel\del\deld={1\over 5040} \\[2mm]\hskip-1cm
I_1^{B_2}=\2i\del|_\t\ \del|_\s\ \del=-{17\over 5040}&\hskip1cm
I_2^{B_2}=\2i\del|_\t(\ddel|_\s)^2\del={1\over 1260} \\[2mm]\hskip-1cm
I_3^{B_2}=\2i(\ddel|_\t)^2(\ddel|_\s)^2\del=-{1\over 4032}&\hskip1cm
I_1^{B_3}=\2i\del=-{1\over 12}\\[2mm]\hskip-1cm
I_2^{B_3}=\2i\del|_\t\ \del={1\over 60} &\hskip1cm
I_3^{B_3}=\2i(\ddel|_\t)^2\del=-{1\over 240}
\end{array}
\EN

\vfill \eject 

\noindent $\bullet$\ \underline{$\langle S_4 S_6 \rangle_c $}
\bea
\langle S_4 S_6 \rangle_c &=&
-{\beta^3\over 5!}\biggl\{
(4K_6+2K_7-8K_8+5K_{11})(I_2^{C_1}+I_3^{C_1}-2I_4^{C_1})\nonumber\\
&&+(-2K_4+2K_5+K_{10}+3K_{12})(I_1^{C_2}+I_2^{C_2}-2I_3^{C_2})\nonumber\\
&&+(2K_4-2K_5+3K_{10}-K_{12})(I_4^{C_2}+I_5^{C_2}-2I_6^{C_2})\nonumber\\
&&+(-4K_4+4K_5-6K_{10}+2K_{12})(I_7^{C_2}+I_8^{C_2}-I_9^{C_2}
-I_{10}^{C_2})\nonumber\\
&&+(4K_4-4K_5+2K_{10}-4K_{12})(I_{11}^{C_2}+I_{12}^{C_2}-2I_{13}^{C_2})\\
&&+{4\over 9}\biggl[
(3K_6-7K_7-2K_8)(I_2^{C_1}+I_3^{C_1}-2I_4^{C_1})\nonumber\\
&&+(4K_5+6K_6)(I_1^{C_2}+I_2^{C_2}-2I_3^{C_2})
+(2K_4+3K_6)(I_4^{C_2}+I_5^{C_2}-2I_6^{C_2})\nonumber\\
&&+(-4K_4-8K_5-6K_{10})(I_7^{C_2}+I_8^{C_2}-I_9^{C_2}
-I_{10}^{C_2})\nonumber\\
&&+(2K_4-4K_5-3K_6)(I_{11}^{C_2}+I_{12}^{C_2}-2I_{13}^{C_2})
\biggr]\biggr\}
-{\beta^3\over 6}R^{mn}\nabla_n\nabla_mV
(I_{1}^{C_3}+I_{2}^{C_3}-2I_{3}^{C_3})
\nonumber
\ena

Integrals:
\EQ
\begin{array}{ll}
\hskip-1cm
I_1^{C_1}=\2i (\overline{\ddeld})|_\t\ \del^2 (\deld)^2={1\over 420}
&\hskip1cm
I_2^{C_1}=\2i \del|_\t\ (\ddel)^2 (\deld)^2=-{1\over 420}\\[2mm]
\hskip-1cm
I_3^{C_1}=\2i \del|_\t\ \del^2 (\ddeld{}^2-\dddel^2)={1\over 280}
&\hskip1cm
I_4^{C_1}=\2i \del|_\t\ \del\ \ddel\ \ddeld\ \deld={1\over 1680}\\[2mm]
\hskip-1cm
I_5^{C_1}=\2i \ddel|_\t\ \del\ \ddel(\deld)^2=0
&\hskip1cm
I_6^{C_1}=\2i \ddel|_\t\ \del^2\ \ddeld\ \deld=-{1\over 840}\\[2mm]
\hskip-1cm
I_1^{C_2}=\2i (\del\ \overline{\ddeld})|_\t\ (\overline{\ddeld})|_\s\  
\del^2=-{1\over 420}
&\hskip1cm
I_2^{C_2}=\2i (\del\ \overline{\ddeld})|_\t\ \del|_\s\  
(\deld)^2={1\over 210}\\[2mm]
\hskip-1cm
I_3^{C_2}=\2i (\del\ \overline{\ddeld})|_\t\ \ddel|_\s\  
\del\deld={1\over 840}
&\hskip1cm
I_4^{C_2}=\2i (\del^2)|_\t\ (\overline{\ddeld})|_\s\ 
(\ddel)^2={1\over 252}\\[2mm]
\hskip-1cm
I_5^{C_2}=\2i (\del^2)|_\t\ \del|_\s\  
(\ddeld{}^2-\dddel^2)={11\over 1260}
&\hskip1cm
I_6^{C_2}=\2i (\del^2)|_\t\ \ddel|_\s\ 
\ddel\ \ddeld=-{1\over 504}\\[2mm]
\hskip-1cm
I_7^{C_2}=\2i (\del\ \ddel)|_\t\ (\overline{\ddeld})|_\s  
\ddel\del={1\over 1260}
&\hskip1cm
I_8^{C_2}=\2i (\del\ \ddel)|_\t\ \del|_\s\ 
\deld\ \ddeld=-{1\over 630}\\[2mm]
\hskip-1cm
I_9^{C_2}=\2i (\del\ \ddel)|_\t\ \ddel|_\s\  
\del\ \ddeld=-{1\over 1008}
&\hskip1cm
I_{10}^{C_2}=\2i (\del\ \ddel)|_\t\ \ddel|_\s\ 
\ddel \deld={1\over 5040}\\[2mm]
\hskip-1cm
I_{11}^{C_2}=\2i (\ddel^2)|_\t\ (\overline{\ddeld})|_\s\ 
\del^2={1\over 2520}
&\hskip1cm
I_{12}^{C_2}=\2i (\ddel^2)|_\t\ \del|_\s\ 
(\deld)^2=-{1\over 1260}\\[2mm]
\hskip-1cm
I_{13}^{C_2}=\2i (\ddel^2)|_\t\ \ddel|_\s\ 
\del\deld=-{1\over 5040}
&\hskip1cm
I_{1}^{C_3}
=\2i  (\overline{\ddeld})|_\t\ \del^2={1\over 90}\\[2mm]
\hskip-1cm
I_{2}^{C_3}=\2i  \del|_\t\ \ddel^2=-{1\over 45} &\hskip1cm
I_{3}^{C_3}=\2i  \ddel|_\t\ \ddel=-{1\over 180}
\end{array}
\EN

\vskip 1cm

\noindent $\bullet$\ \underline{$\langle S_8 \rangle $}
\bea
 \langle S_8 \rangle &=& {\beta^3 \over 7!}(I^D_1-I^D_2) 
 \biggl[ 4( 2 K_4 + 9K_6-7K_7 -2K_8 )
+ {55\over 2} 
(11 K_{13} -10 K_{14} +3 K_{15} )
\nonumber \\ && 
+34( 2 K_4 -2K_5 +4K_6 +2K_7 -8K_8 +3 K_{10} +5K_{11}-K_{12})
\nonumber \\ && 
+ 5 (12 K_4 
-12K_5 +4K_6 +2K_7 -8K_8 +6 K_{10} +2K_{11}-16K_{12}+14K_{13}-20K_{14} 
\nonumber \\ &&
-5 K_{16} +9K_{17})\biggr]
-{\beta^3 \over 24} I^D_3 
( 2 R^{mn}\nabla_m\nabla_n V +
\nabla^m R \nabla_m V - 3 \nabla^4 V )
\eea

Integrals:
\EQ
\begin{array}{lll}
I^D_1 =\int \del^3\:{(\ddeld +\deldd)}= -{1\over 140}; & \
I^D_2 = \int \del^2\:{\deld}^2 = {1\over 840} ; & \
I^D_3 = \int \del^2 = {1\over 30}. 
\end{array}
\EN
}


\end{document}